\pdfoutput=1
\documentclass{pnastwo}
\usepackage{graphicx}

\begin{document}

\title{Scaling metagenome sequence assembly with probabilistic de Bruijn
graphs}

\author{
Jason Pell
\affil{1}{Computer Science and Engineering, Michigan State University, East Lansing, MI, United States}
\and
Arend Hintze
\affil{1}{}
\and
Rosangela Canino-Koning
\affil{1}{}
\and
Adina Howe
\affil{2}{Microbiology and Molecular Genetics, Michigan State University, East Lansing, MI, United States}
\and
James M. Tiedje
\affil{2}{}
\affil{3}{Crop and Soil Sciences, Michigan State University, East Lansing, MI, United States}
\and
C. Titus Brown
\affil{1}{}
\affil{2}{}
\footnote{Corresponding author: ctb@msu.edu}
}

\maketitle


\begin{article}

\begin{abstract}

Deep sequencing has enabled the investigation of a wide range of
environmental microbial ecosystems, but the high memory requirements
for {\em de novo} assembly of short-read shotgun sequencing data from these
complex populations are an increasingly large practical barrier.  Here
we introduce a memory-efficient graph representation with which we can
analyze the k-mer connectivity of metagenomic samples.  The graph
representation is based on a probabilistic data structure, a Bloom
filter, that allows us to efficiently store assembly graphs in as
little as 4 bits per k-mer, albeit inexactly.  We show that this data structure
accurately represents DNA assembly graphs in low memory.  We apply
this data structure to the problem of partitioning assembly graphs
into components as a prelude to assembly, and show that this reduces
the overall memory requirements for {\em de novo} assembly of
metagenomes.  On one soil metagenome assembly, this approach achieves
a nearly 40-fold decrease in the maximum memory requirements for
assembly.  This probabilistic graph representation is
a significant theoretical advance in storing assembly graphs and also yields
immediate leverage on metagenomic assembly.

\end{abstract}

\keywords{metagenomics | de novo assembly | de Bruijn graphs | k-mers}

\section{Introduction}

{\em De novo} assembly of shotgun sequencing reads into longer contiguous
sequences plays an important role in virtually all genomic research
\cite{pubmed19482960}.  However, current computational methods for
sequence assembly do not scale well to the volume of sequencing data
now readily available from next-generation sequencing machines
\cite{pubmed19482960,pubmed22147368}.  In particular, the deep
sequencing required to sample complex microbial environments easily
results in data sets that surpass the working memory of available
computers \cite{metahit,rumen}.

Deep sequencing and assembly of short reads is particularly important
for the sequencing and analysis of complex microbial ecosystems, which
can contain millions of different microbial species
\cite{pubmed20195499,pubmed16123304}.  These ecosystems mediate
important biogeochemical processes but are still poorly understood at
a molecular level, in large part because they consist of many microbes
that cannot be cultured or studied individually in the lab
\cite{pubmed20195499,nrcbook}.  Ensemble sequencing (``metagenomics'') of
these complex environments is one of the few ways to render them
accessible, and has resulted in substantial early progress in
understanding the microbial composition and function of the ocean,
human gut, cow rumen, and permafrost soil
\cite{metahit,rumen,sargasso,permafrost}.  However, as sequencing
capacity grows, the assembly of sequences from these complex samples
has become increasingly computationally challenging.  Current methods
for short-read assembly rely on inexact data reduction in which reads
from low-abundance organisms are discarded, biasing analyses towards
high-abundance organisms \cite{metahit,rumen,permafrost}.


The predominant assembly formalism applied to short-read sequencing
data sets is a de Bruijn graph
\cite{pubmed11504945,pubmed20211242,pubmed22068540}.  In a de Bruijn
graph approach, sequencing reads are decomposed into fixed-length
words, or k-mers, and used to build a connectivity graph.  This graph
is then traversed to determine contiguous sequences
\cite{pubmed22068540}.  Because de Bruijn graphs store only k-mers,
memory usage scales with the number of unique k-mers in the data set
rather than the number of reads \cite{succinct,pubmed22068540}.  Thus
human genomes can be assembled in less than 512GB of system memory
\cite{pmid21187386}.  For more complex samples such as soil
metagenomes, which may possess millions or more species, terabytes of
memory would be required to store the graph.  Moreover, the wide
variation in species abundance limits the utility of standard
memory-reduction practices such as abundance-based error-correction
\cite{pubmed21114842}.

In this work, we describe a simple probabilistic representation for
storing de Bruijn graphs in memory, based on Bloom filters
\cite{bloom}.  Bloom filters are fixed-memory probabilistic data
structures for storing sparse sets; essentially hash tables without
collision detection, set membership queries on Bloom filters can yield
false positives but not false negatives. While Bloom filters have been used
in bioinformatics software tools in the past, they have not been used
for storing assembly
graphs\cite{bloomref1,bloomref2,bloomref3,bloomref4}. We show that
this probabilistic graph representation more efficiently stores de
Bruijn graphs than any possible exact representation, for a wide range
of useful parameters.  We also demonstrate that it can be used to
store and traverse actual DNA de Bruijn graphs with a 20- to 40-fold
decrease in memory usage over two common de Bruijn graph-based
assemblers, Velvet and ABySS~\cite{velvet,abyss}. We relate changes in
local and global graph connectivity to the false positive rate of the
underlying Bloom filters, and show that the graph's global structure
is accurate for false positive rates of 15\% or lower, corresponding
to a lower memory limit of approximately 4 bits per graph node.

We apply this graph representation to reduce the memory needed to
assemble a soil metagenome sample, through the use of read
partitioning.  Partitioning separates a de Bruijn graph up into
disconnected graph components; these components can be used to
subdivide sequencing reads into disconnected subsets that can be
assembled separately.  This exploits a convenient biological feature
of metagenomic samples: they contain many microbes that should not
assemble together.  Graph partitioning has been used to improve the
quality of metagenome and transcriptome assemblies by adapting
assembly parameters to local coverage of the
graph~\cite{metavelvet,pubmed21685107,trinity}.  However, to our
knowledge, partitioning has not been applied to scaling metagenome
assembly. By applying the probabilistic de Bruijn graph representation
to the problem of partitioning, we achieve a dramatic decrease of
nearly 40-fold in the memory required for assembly of a soil
metagenome.


\section{Results}

\subsection{Bloom filters can store de Bruijn graphs}

Given a set of short DNA sequences, or reads, we first break down each
read into a set of overlapping k-mers.  We then store each k-mer in a
Bloom filter, a probabilistic data structure for storing elements from
sparse data sets (see \emph{Methods} for implementation details).
Each k-mer serves as a vertex in a graph, with an edge between two
vertices $N_1$ and $N_2$ if and only if $N_1$ and $N_2$ share a
(k-1)-mer that is a prefix of $N_1$ and a postfix of $N_2$, or vice
versa.  This edge is not stored
explicitly, which can lead to false connections when two reads abut
but do not overlap; these false connections manifest as false positives,
discussed in detail below.

Thus each k-mer has up to 8 edges connecting to 8 neighboring k-mers,
which can be determined by simply building all possible 1-base
extensions and testing for their presence in the Bloom filter.  In
doing so, we implicitly treat the graph as a simple graph as opposed
to a multigraph, which means that there can be no self-loops or
parallel edges between vertices/k-mers.  By relying on Bloom filters,
the size of the data structure is fixed: no extra memory is used as
additional data is added.

This graph structure is effectively {\em compressible} because one can
choose a larger or smaller size for the underlying Bloom filters; for
a fixed number of entries, a larger Bloom filter has lower occupancy
and produces correspondingly fewer false positives, while a smaller
Bloom filter has higher occupancy and produces more false
positives. In exchange for memory, we can store k-mer nodes more or
less accurately: for example, for a false positive rate of 15\%, at
which 1 in 6 random k-mers tested would be falsely considered present,
each real k-mer can be stored in under 4 bits of memory (see
Table 1).  While there are many false k-mers, they only matter if they
connect to a real k-mer.

The false positive rate inherent in Bloom filters thus raises one concern
for graph storage:
in contrast to an exact graph storage, there is
a chance that a k-mer will be adjacent to a false positive k-mer.
That is, a k-mer may connect to another k-mer that does not actually
exist in the original dataset but nonetheless registers as present,
due to the probabilistic nature of the Bloom filter.  As the memory
per real k-mer is decreased, false positive vertices and edges are
gained, so compressing the graph results in a more tightly
interconnected graph.  If the false positive rate is too high, the
graph structure will be dominated by false connectivity -- but what
rate is ``too high''?  We study this key question in detail below.

\subsection{False positives cause local elaboration of graph structure}

\begin{figure}
\includegraphics[width=2in]{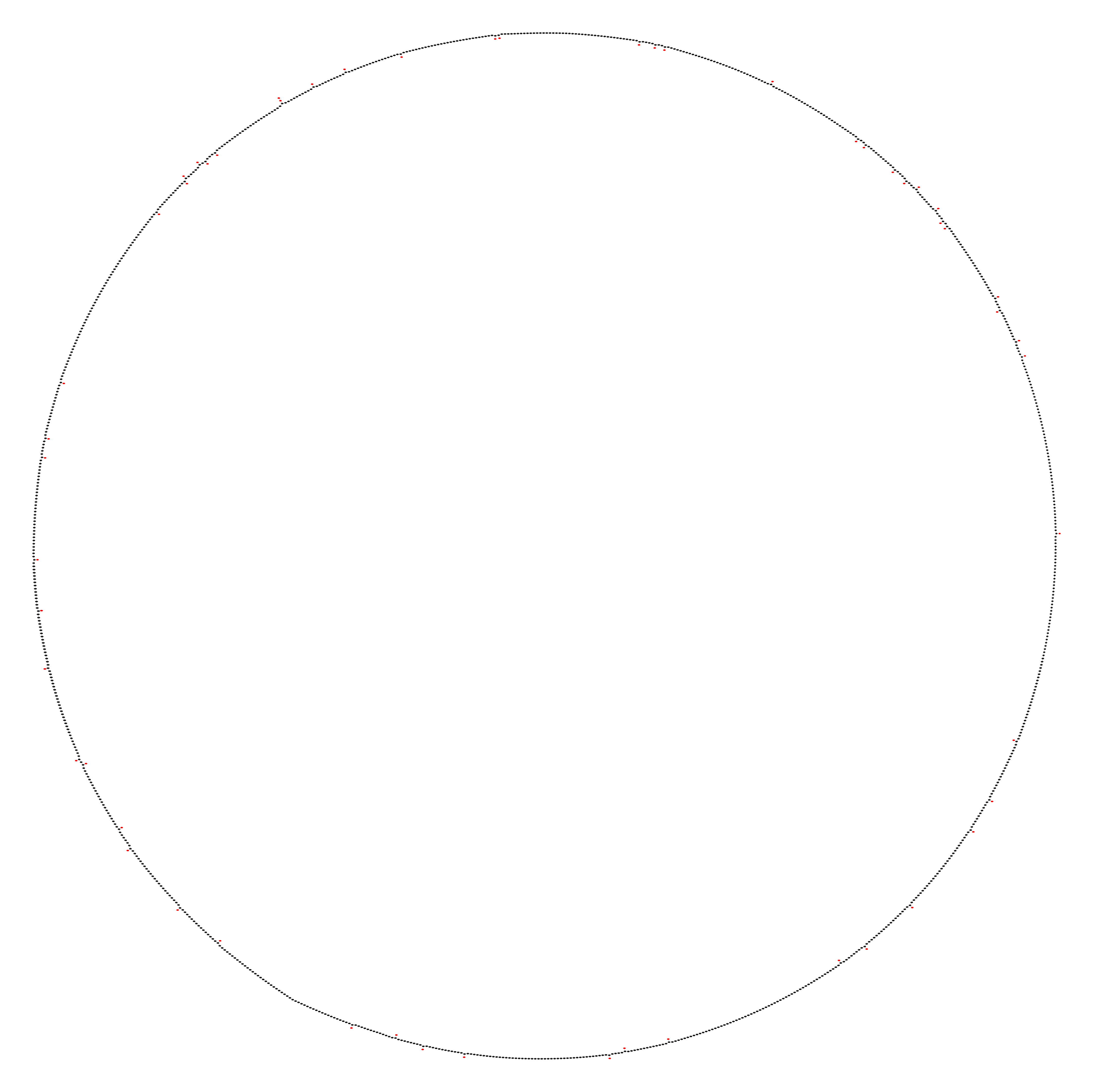}
\includegraphics[width=2in]{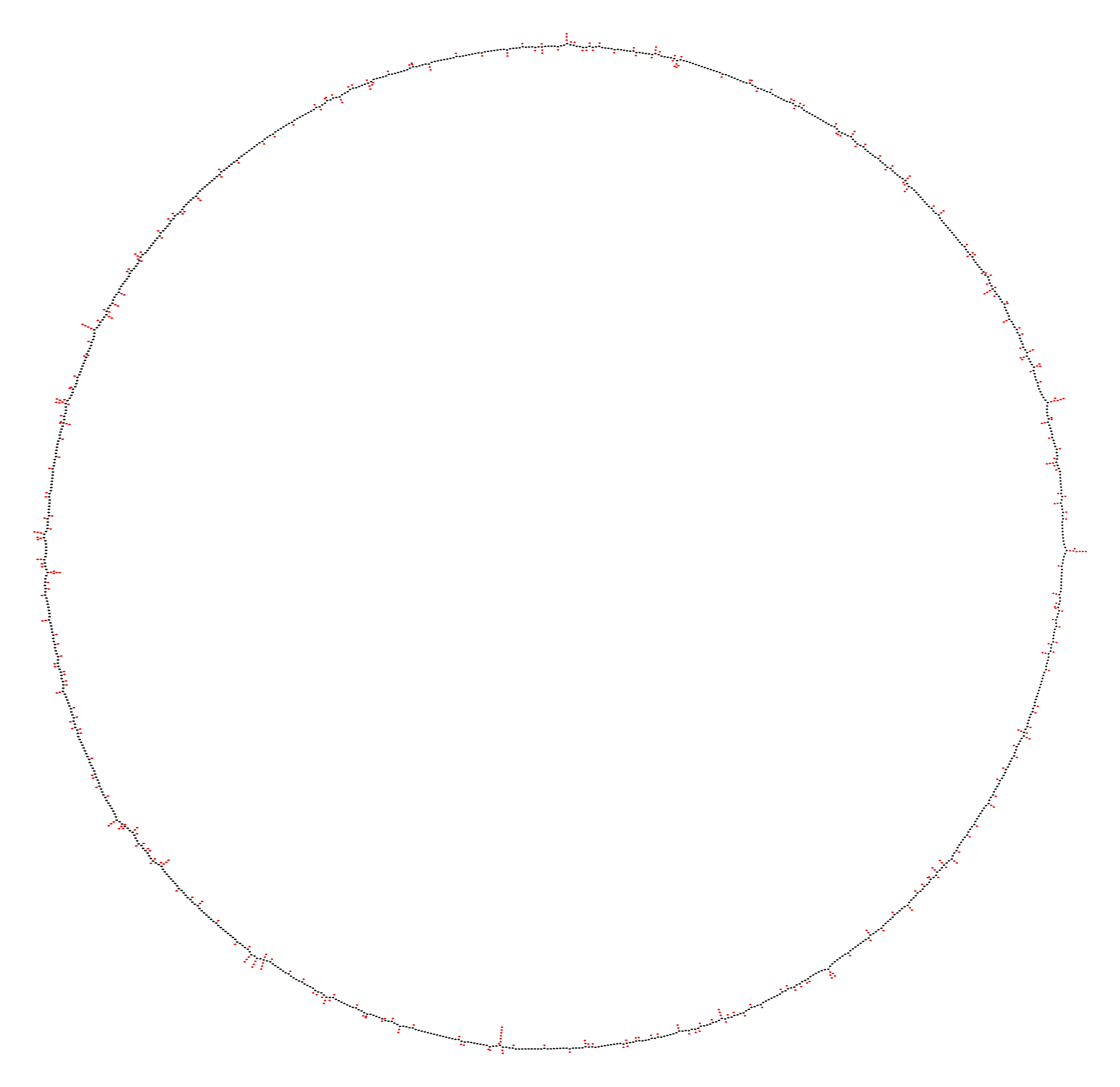}
\includegraphics[width=2in]{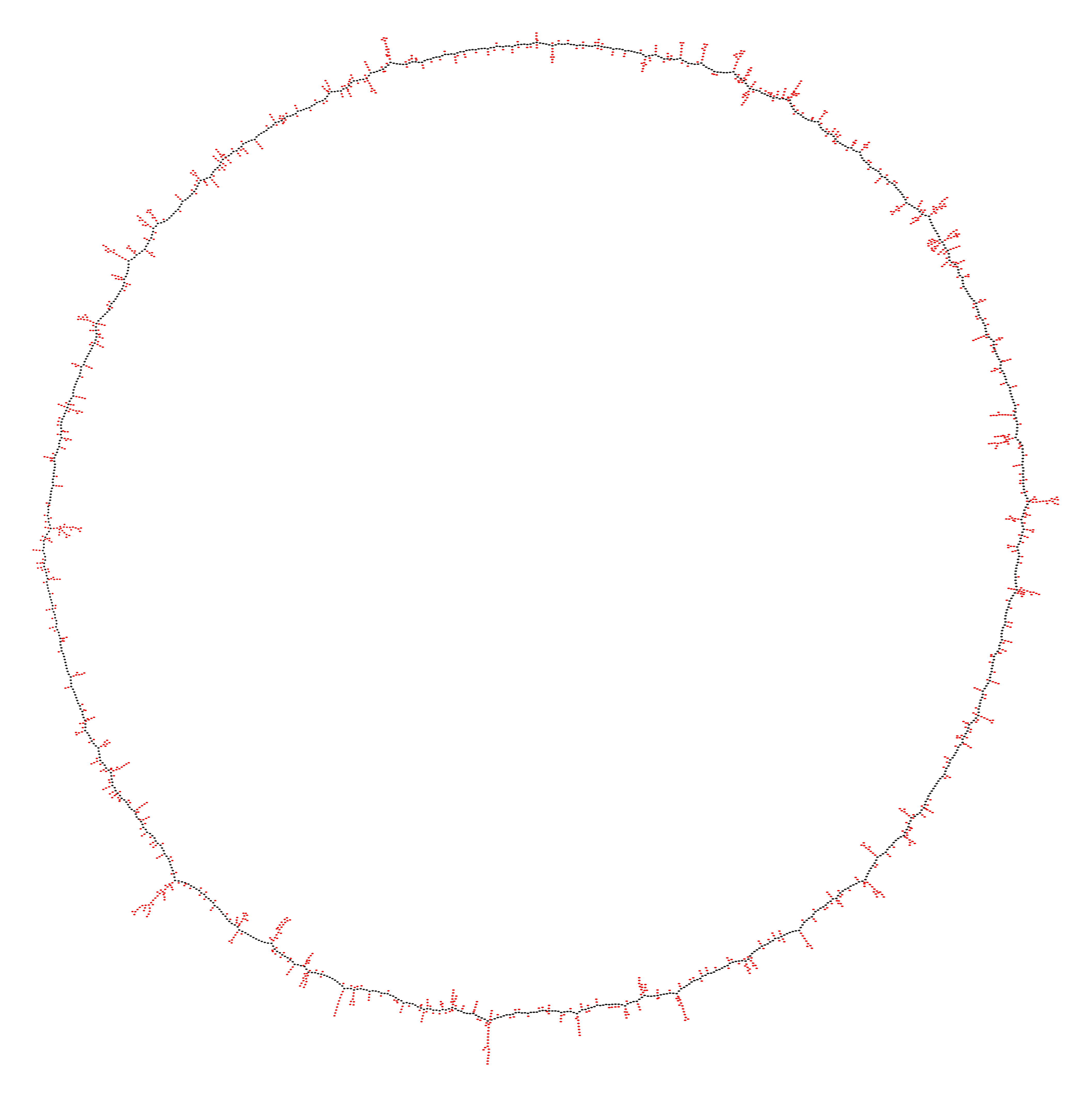}
\includegraphics[width=2in]{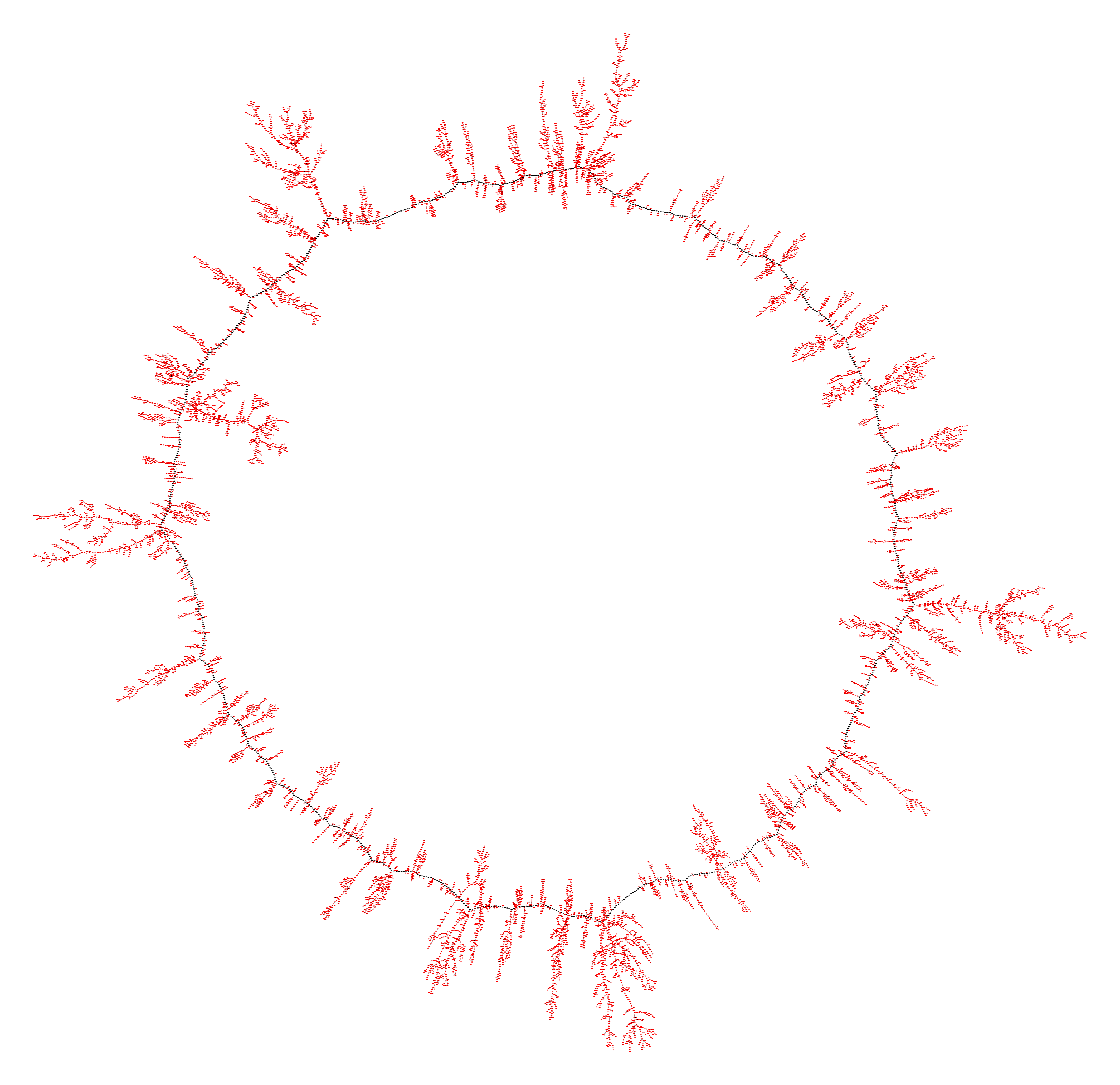}

\caption{Graph visualizations demonstrating the decreasing fidelity of
  graph structure with increasing false positive rate. Erroneous k-mers are 
  colored red and k-mers corresponding to the original generated sequence 
  (1,000 31-mers generated by a 
  1,031 bp circular chromosome) 
  are black. From top left
  to bottom right, the false positive rates are 0.01, 0.05, 0.10, and
  0.15.  Shortcuts ``across'' the graph are not created.}

\label{fig:circles}
\end{figure}

Erroneous neighbors created by false positives can alter the graph
structure.  To better understand this effect, we generated a random
1,031bp circular sequence and visualized the effect of different false
positive rates.  After storing this single sequence in compressible
graphs using $k=31$ with four different false positive rates
($p_f$=0.01, 0.05, 0.10, and 0.15), we explored the graph using
breadth-first search beginning at the first 31-mer.  The graphs in
Figure \ref{fig:circles} illustrate how the false positive k-mers 
connected to the original k-mers (from the 1,031bp sequence) 
elaborate with the false positive rate while the overall circular
graph structure remains, with no erroneous shortcuts between k-mers
that are present in the original sequence.  It is visually apparent
that even a high false positive rate of 15\% does not systematically
and erroneously connect distant k-mers.


\subsection{False long-range connectivity is a nonlinear function of the false positive rate}

\begin{figure}
\center{\includegraphics[width=5in]{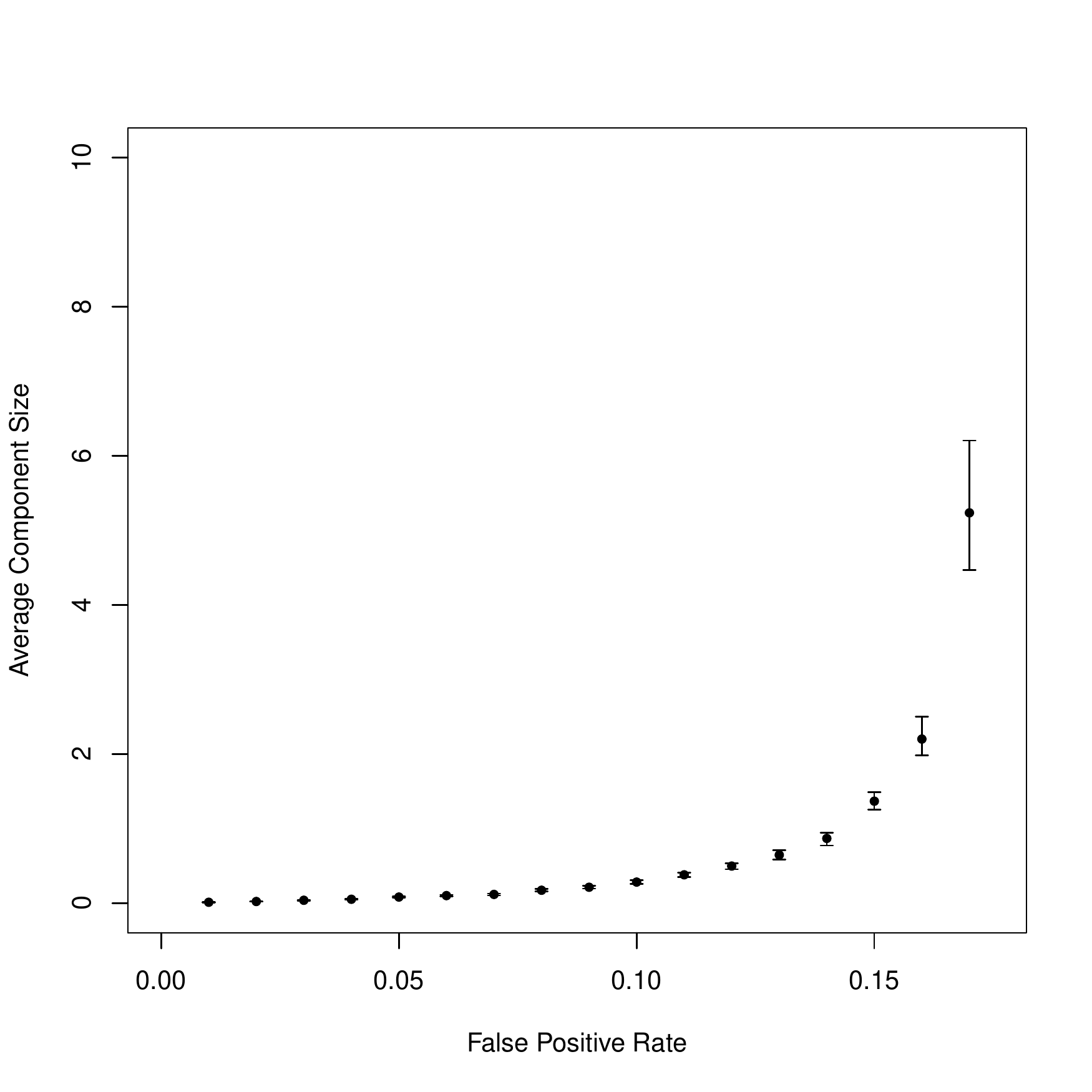}}
\caption{Average component size versus false positive rate. The average 
component size sharply increases as the false positive 
rate approaches the percolation threshold.
}
\label{fig:clustersize}
\end{figure}

To explore the point at which our data structure systematically
engenders false long-range connections, we inserted random k-mers into
Bloom filters with increasing false positive rates.  These k-mers
connect to other k-mers to form graph components that increase in size
with the false positive rate.  We then calculated the average
component size in the graph for each false positive rate ($n=10000$)
and used percentile bootstrap to obtain estimates within a 95\%
confidence interval. Figure \ref{fig:clustersize} demonstrates that
the average component size rapidly increases as a specific threshold
is approached, which appears to be at a false positive rate near 0.18
for k=31. Beyond 0.18, the components begin to join together into
larger components.

As the false positive rate increases, we observe a sudden transition
from many small components to fewer, larger components created by
erroneous connections between the ``true'' components (Figure
\ref{fig:clustersize}).  In contrast to the linear increase in the
local neighborhood structure as the false positive rate increases
linearly, the change in global graph structure is abrupt as previously
disconnected components join together.  This rapid change resembles a
geometric phase transition, which for graphs can be discussed in terms
of percolation theory~\cite{staufferintro}. We can map our problem to
site percolation by considering a probability $p$ that a particular
k-mer is present, or ``on''. (This is in contrast to bond percolation
where $p$ represents the probability of a particular edge being
present.) As long as the false positive rate is below the percolation
threshold $p_\theta$ (i.e. in the subcritical phase), we would predict
that the graph is not highly connected by false positives.


Percolation thresholds for finite graphs can be estimated by finding
where the component size distribution transitions from linear to
quadratic in form \cite{stauffer1979scaling}.  Using the calculation
method described in \emph{Methods}, we found the site percolation
threshold for DNA de Bruijn graphs to be $p_\theta = 0.183 \pm 0.001$
for k between 5 and 12.  Although we only tested within this limited
range of k, the percolation threshold appears to be independent of
different $k$ (see Figure~S1).  Thus, as long as the false positive
rate is below $0.183$, we predict that truly disconnected components
in the graph are unlikely to connect to one another erroneously, that
is, due to errors introduced by the probabilistic representation.

\subsection{Large-scale graph structure is retained up to the percolation threshold}

The results from component size analysis and the percolation threshold
estimation suggest that global connectivity of the graph is unlikely
to change below a false positive rate of 18\%.  Do we see this invariance
in global connectivity in other graph measures?

To assess global connectivity, we employed the diameter metric in
graph theory, the length of the ``longest shortest'' path between any
two vertices~\cite{bondy2008graph}.  If shorter paths between real
k-mers were being
systematically generated due to false positives, we would expect the
diameter of components to decrease as the false positive rate
increased.  We randomly generated 58bp long circular chromosomes (50bp
read with the first 8-mer appended to the end of the string) to
construct components containing 50 8-mers and calculated the diameter
at different false positive rates. We kept $k$ low because we needed
to be able to exhaustively explore the graph even beyond the
percolation threshold, which is computationally prohibitive to do at
higher k values. Furthermore, larger circular chromosomes would be
more likely to erroneously connect at a fixed $k$, but due to the
relatively low number of possible 8-mers, we had to keep the
chromosomes small.  We only considered paths between two real k-mers
in the dataset.

\begin{figure}
\center{\includegraphics[width=5in]{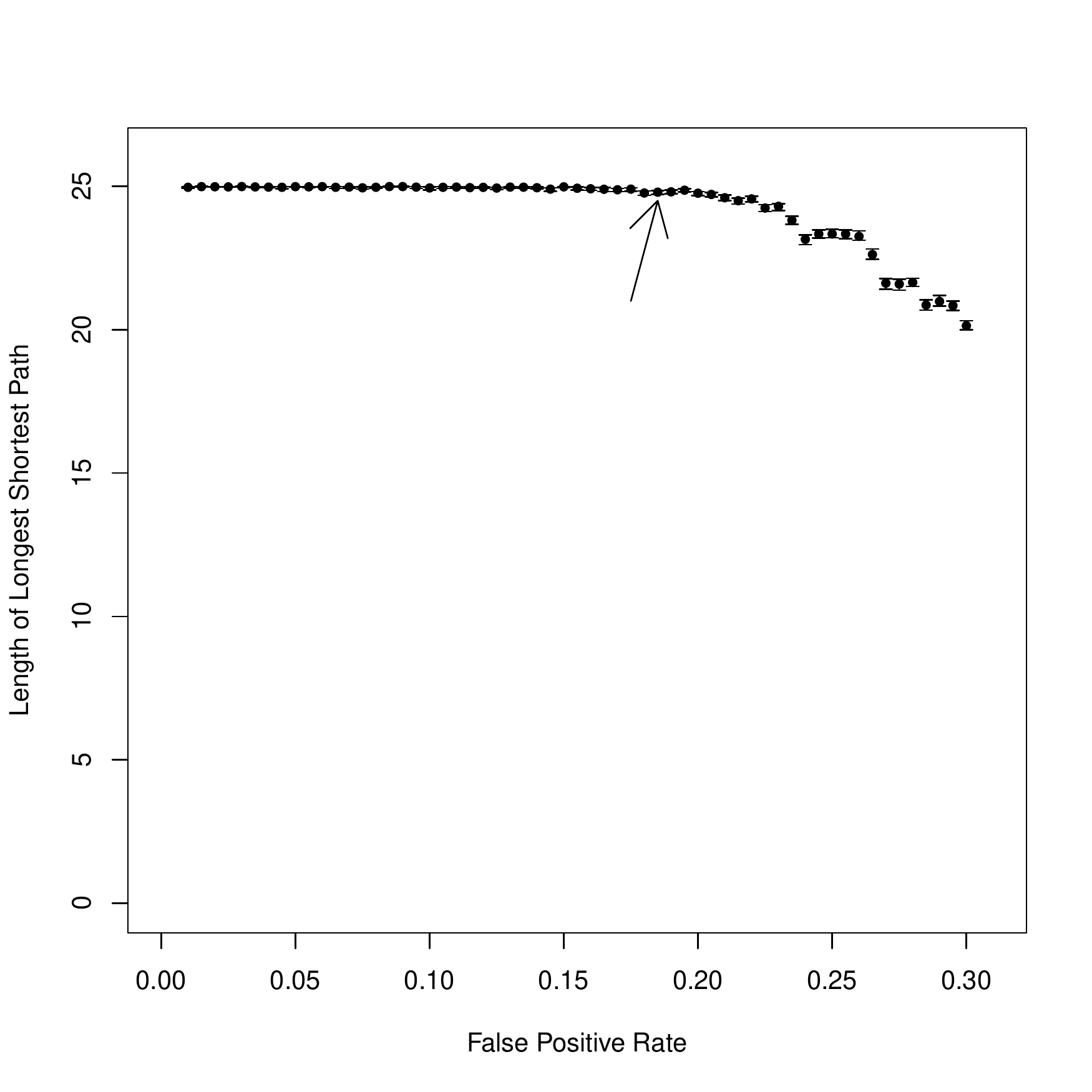}}

\caption{The diameter of randomly generated 58bp long circular
  chromosomes in 8-mer (i.e. a cycle of 50 8-mers) space remains 
constant for false
  positive rates up through 18.3\%. Only real (non-error) k-mers are considered for
  starting and ending points.}
\label{fig:diam}
\end{figure}

At each false positive rate, we ran the simulation 500 times and
estimated the mean within a 95\% confidence interval using percentile
bootstrap. As Figure~\ref{fig:diam} shows, erroneous connections
between pairs of real k-mers are rare below a false positive rate of
20\%.  For false positive rates above this threshold, spurious
connections between real k-mers are created, which lowers the
diameter.  Thus, the larger scale graph structure is retained up
through $p = 0.183$, as suggested by the component size analysis and
percolation results.  This demonstrates that as long as the k-mer
space is only sparsely occupied by false positives, long ``bridges''
between distant k-mers do not appear spontaneously.

\subsection{Erroneous k-mers from sequencing eclipse graph false positives}


It is important to compare the errors from false positives in the de
Bruijn graph representation with errors from real data.  In
particular, real data from massively parallel sequencers will contain
base calling errors.  In de Bruijn graph-based assemblers, these
sequencing errors add to the graph complexity and make it more
difficult to find high-quality paths for generating long, accurate
contigs. Since our approach also generates false positives, we wanted
to compare the error rate from the Bloom filter graph with
experimental errors generated by sequencing (Table
2). We used the \emph{E. coli} K-12 MG1665 genome to
compare various graph invariants between an Illumina dataset generated
from the same strain (see \emph{Methods}), an exact representation of
the genome, and inexact representations with different false positive
rates.

For these comparisons, we used a $k$ value of 17, for which we can
store graphs exactly,
i.e. we have no false
positives because we can store $4^{17}$ entries precisely in 2GB of
system memory. This is equivalent to a Bloom filter with one hash
table and a 0\% false positive rate.  We found a total of 50,605
17-mers in the exact representation that were not part of a simple
line, i.e. had more than two neighbors (degree $>$ 2). As the false
positive rate increased, the number of these 17-mers increased in the
expected linear fashion.
Furthermore, the number of real 17-mers, those that are not false
positives, comprise the majority of the graph.
(As above, we only counted false positive k-mers
that are transitively connected to at least one real k-mer.)

In contrast, when we examined an exact representation of an Illumina
dataset, only 9.9\% of the k-mers in the graph truly exist in the
reference genome.   The number of
17-mers with more than 2 neighbors in the sequencing reads is higher than for the exact
representation of the genome, which demonstrates that sequencing
errors add to the complexity of the graph. Overall, the errors
demonstrated by sequencers dwarf the errors caused by the inexact
graph representation at a reasonable false positive rate.

When we assemble this data set with the Velvet and ABySS assemblers at
k=31, Velvet requires 3.7GB to assemble the data set, while ABySS
requires 1.6GB; this memory usage is dominated by the graph storage
\cite{zerbinothesis}. Thus the Bloom filter approach stores graphs 30
or more times more efficiently than either program, even with a low
false positive rate of 1\%.  While this direct comparison can not be
made fairly -- assemblers store the graph as well as k-mer abundances
and other information -- it does suggest that there are opportunities
for decreasing memory usage with the probabilistic graph representation.

\subsection{Sequences can be accurately partitioned by graph connectivity}

Can we use this low-memory graph representation to find and separate
components in de Bruijn graphs?  The primary concern is that false
positive nodes or edges would connect components, but the diameter
results suggest that components are unlikely to connect below a 20\%
false positive rate.  To verify this, we analyzed a simulated dataset
of 1,000 randomly generated sequences of length 10,000 bp.  Using
$k=31$, we partitioned the data across many different false positive
rates, using the procedure described in \emph{Methods}. As predicted,
the resulting number of partitions did not vary across the false
positive rates while $f_p \le 0.15$ (Figure~S2).

We then applied partitioning to a considerably larger bulk soil
metagenome (``MSB2'') containing 35 million 75 bp long reads generated
from an Illumina GAII sequencer.  We calculated the number of unique
31-mers present in the data set to be 1.35 billion. Then, for each of
several false positive rates (see Table 
3) we loaded
the reads into a graph, eliminated components containing fewer than
200 unique k-mers, and partitioned the reads into separate files based
on graph connectivity.

Once we obtained the partition sets, we individually assembled each
set of partitions using ABySS, as well as the entire (unpartitioned)
data set, retaining contigs longer than 500 bp.  The resulting
assemblies were all identical, containing 1,444 contigs with a total
assembled sequence of 1.07 megabases.  The unpartitioned data set
required 33GB to assemble with ABySS, while the data set could be
partitioned in under 1 GB with a 30-fold decrease in maximum memory usage
(Table 3).
Moreover, despite this dramatic decrease in the memory required to assemble
the data set, the assembly results are identical.


\section{Discussion}


\subsection{Bloom filters can be used to efficiently store de Bruijn graphs}

\begin{figure}
\center{\includegraphics[width=6in]{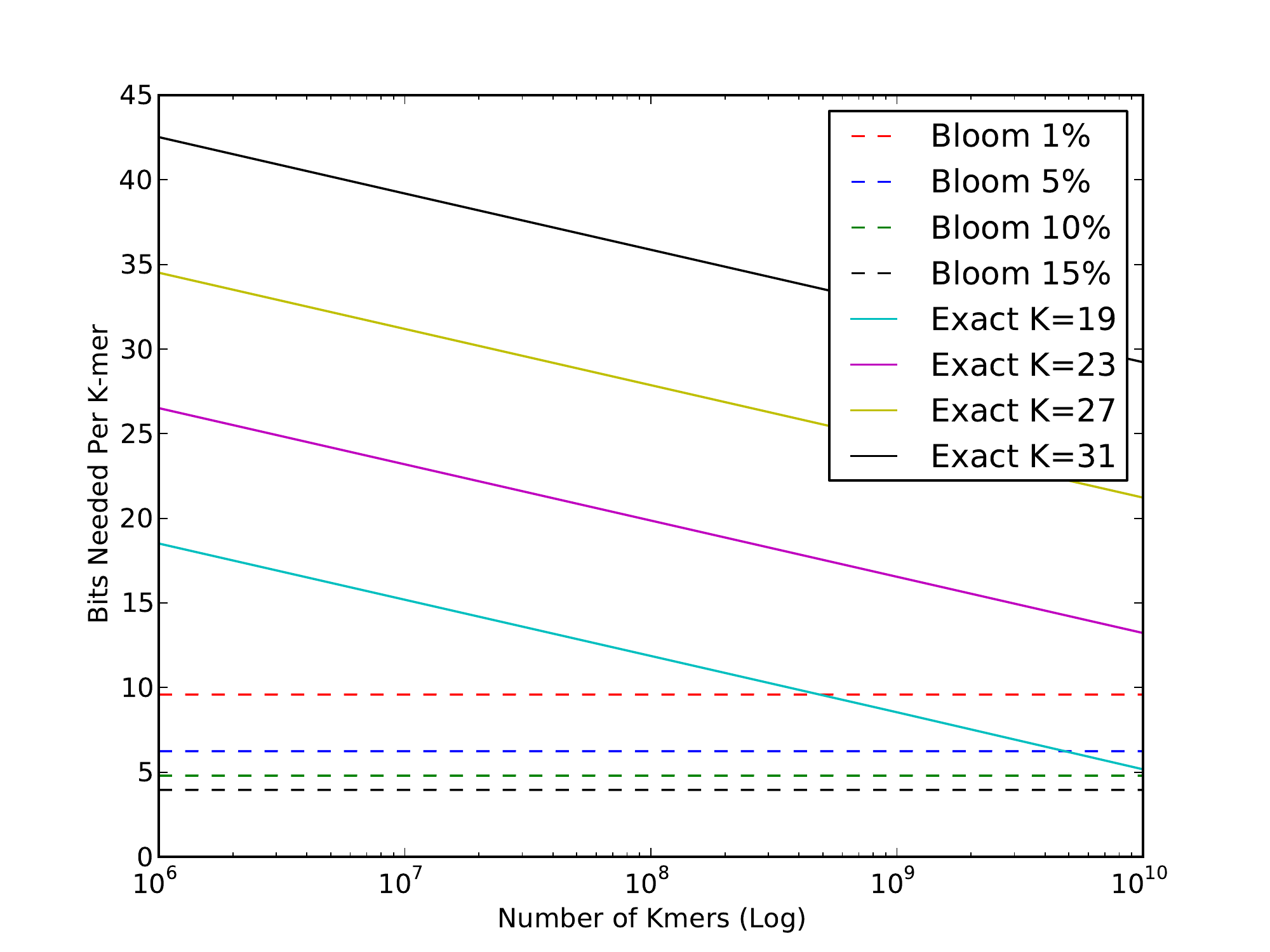}}

\caption{Comparison between Bloom filters at different false positive 
rates with the information-theoretic lossless lower bound at different 
k values. Bloom filters are k independent and are more efficient than 
any lossless data structure for higher k due to greater sparseness in 
k-mers inserted compared to all possible k-mers.}

\label{fig:membound}
\end{figure}

The use of Bloom filters to store a de Bruijn graph is straightforward
and memory efficient.  The expected false positive rate can be tuned
based on desired memory usage, yielding a wide range of possible
storage efficiencies (Table 1).
Since memory usage
is k independent in Bloom filters, it is more efficient than the
theoretical lower-bound for a lossless exact representation when the
number of k-mers inserted in the graph is sparsely populated, which is
dependent on k (Figure \ref{fig:membound}; see \cite{succinct} for
details on lower-bound memory usage for an exact representation).

Even for low false positive rates such as 1\%, this is still an
efficient graph representation, with significant improvements in both
theoretical memory usage (Figure \ref{fig:membound}) and actual memory 
usage compared
to two existing assemblers, Velvet and ABySS (Table
2). We can store k-mers in this data structure with a
much smaller set of ``erroneous'' k-mers than those generated by
sequencing errors, and the Bloom filter false positive rates have less
of an effect on branching graph structure than do sequencing errors.
In addition, the false positives engendered by the Bloom filters are
uncorrelated with the original sequence, unlike single-base sequencing
errors which resemble the real sequence.

Using a probabilistic graph representation with false positive nodes
and edges raises the specter of systematic graph artifacts resulting
from the false positives.  For partitioning, our primary concern was
that false positives would incorrectly connect components, rendering
partitioning ineffective.  The results from percolation analysis,
diameter calculations, and partitioning of simulated and real data
demonstrate that below the calculated percolation threshold there is
no significant false connectivity.  As long as the false positive rate
is below 18.3\%, long false paths are not spontaneously created and
the large scale graph properties do not change.  Above this rate, the
global graph structure slowly degrades.

\subsection{Partitioning works on real data sets}

Our partitioning results on a real soil metagenome, the MSB2 data set,
demonstrate the utility of partitioning for reducing memory usage.
For this specific data set, we obtained {\em identical} results with a
20-40x decrease in memory (Table 3).
This is
consonant with our results from storing the E. coli genome, in which
we achieved a 30-fold decrease in memory usage over the exact
representation at a false positive rate of 1\%.  While increased
coverage and variation in data set complexity will affect actual
memory usage for other data sets, these results demonstrate that
significant scaling in the memory required for assembly can be
achieved in one real case.

The memory requirements for the partitioning process on the MSB2 data
set are dominated by the memory required to store and explore the
graph; the higher memory usage observed for partitioning at a false
positive rate of 15\% is due to the increase in component size from
local false positives.  Regardless, the memory requirements for
downstream assembly of partitions is driven by the size of the largest
partition, which here is very small (345,000 reads; Table 3)
due to the high diversity of soil and the
concordant low coverage.  The dominant partition size is remarkably
refractory to the graph's false positive rate, increasing by far less
than 1\% for a 15-fold increase in false positives; this shows that
our theoretical and simulated results for component size and diameter
apply to the MSB2 data set as well.



Once partitioned, components can be assembled with parameters chosen
for the coverage and sequence heterogeneity present in each partition.
Moreover, data sets partitioned at a low $k_0$ can be exactly
assembled with any $k \ge k_0$, because overlaps of $k_0-1$ bases
include all overlaps of greater length.  Because the partitions will
generally be much smaller than the total data set (see Table
3), 
they can be quickly assembled with many different
parameters.  This ability to quickly explore many parameters could
result in significant improvement in exploratory metagenome assembly,
where the ``best'' assembly parameters are not known and must be
determined empirically based on many different assemblies.


Combined with the scaling properties of the graph representation,
partitioning with this probabilistic de Bruijn graph representation
offers a way to efficiently apply a partitioning strategy to certain
assembly problems.  While this work focuses on theoretical properties
of the graph representation and analyzes only one real data set, the
results are promising; the next step is to evaluate the approach
on many more real data sets.

\subsection{Concluding thoughts}


Developing efficient and accurate approaches to {\em de novo} assembly
continues to be one of the ``grand challenges'' of bioinformatics
\cite{pubmed22147368}.  Improved metagenome assembly approaches are
particularly important for the investigation of microbial life on
Earth, much of which has not been sampled
\cite{terabasemetag,nrcbook}.  While our appreciation for the role
that microbes play in biogeochemical processes is only growing, we are
increasingly limited by our ability to analyze the data.  For example,
the Earth Microbiome Project is generating petabytes of sequencing
data from complex microbial communities, many of which contain
entirely novel ensembles of microbes; scaling {\em de novo} assembly is a
critical requirement of this investigation \cite{emp2010}.

The probabilistic de Bruijn graph representation presented here has a
number of convenient features for storing and analyzing large assembly
graphs.  First, it is efficient and compressible: for a given data
set, a wide range of false positive rates can be chosen without
impacting the global structure of the graph, allowing graph storage in
as little as 4 bits per k-mer.  Because a higher false positive rate
yields a more elaborate local structure, memory can be traded for
traversal time in e.g. partitioning.  Second, it is a fixed-memory
data structure, with predictable degradation of both local and global
structure as more data is inserted.  For data sets where the number of
unique k-mers is not known in advance, the occupancy of the Bloom
filter can be monitored as data is inserted and directly converted to
an expected false positive rate.  Third, the memory usage is
independent of the k-mer size chosen, making this representation
convenient for exploring properties at many different parameters.  It
also allows the storage and traversal of de Bruijn graphs at multiple
k-mer sizes within a single structure, although we have not yet
explored these properties.  And fourth, it supports memory-efficient
partitioning, an approach that exploits underlying biological features
of the data to divide the data set into disconnected subsets.

Our initial motivation for developing this use of Bloom filters was to
explore partitioning as an approach to scaling metagenome assembly,
but there are many additional uses beyond metagenomics.  Here we
describe exact partitioning of the graph into components, but {\em
  inexact} partitioning has been successfully applied to mRNAseq
assembly \cite{trinity}.  Inexact partitioning, as done by the
Chrysalis component of the Trinity pipeline, uses heuristics to
subdivide the graph for later assembly; the data structure described
in this work can be used for this purpose as well.  More broadly, a
more memory efficient de Bruijn graph representation opens up many
additional opportunities.  While de Bruijn graph approaches are
currently being used primarily for the purposes of assembly, they are
a generally useful formalism for sequence analysis. In particular,
they have been extended to efficient multiple-sequence alignment,
repeat discovery, and detection of local and structural sequence
variation \cite{zerbinothesis,zhang2003dna,price2005novo,cortex}.

We are further exploring the use of probabilistic de Bruijn graphs for
graph-based error correction, graph-based homology search, and
graph-based artifact detection in large sequencing data sets.  The
approach used in partitioning, in which we use properties measured
in the de Bruijn graph to select and extract individual reads based
on e.g. local graph connectivity or sequence identiy, may generalize
to a variety of problems.

\begin{materials}

\section{Genome and Sequence Data}
We used the \emph{E. coli} K-12 MG1655 genome (GenBank: U00096.2) and
two MG1655 Illumina datasets (Short Read Archive accessions SRX000429
and SRX000430) for E. coli analyses.  The MSB2 soil data set is
available as SRA accession SRA050710.1.

\section{Data Structure Implementation}
We implemented a variation on the Bloom filter data structure to store
k-mers in memory. In a classic Bloom filter, multiple hash functions
map bits into a single hash table to add an object or test for the
presence of an object in the set. In our variant, we use multiple
prime-number-sized hash tables, each with a different hash function
corresponding to the modulus of the DNA bitstring representation with
the hash table size; this is a computationally convenient way to
construct hash functions for DNA strings.  The properties
of this implementation are identical to a classical Bloom filter
\cite{bloomsurvey}.




\section{Estimating False Positive Rate For Erroneous Connectivity}
We ran a simulation to find when components in the graph begin to
erroneously connect to one another.  To calculate the false positive
rate $p$ at which this aberrant connectivity occurs, we added random
k-mers, sampled from a uniform GC distribution, to the data structure
and then calculated the occupancy and size of the largest
component. From this we sampled the relative size of the largest
component and the overall component size distribution for each given
occupancy rate.  At the occupancy where a ``giant component'' appears,
this component size distribution should be scale-free
\cite{stauffer1979scaling}.
We then found at what value of $p$ the resulting component size
distribution in logarithmic scale can be better fitted in a linear or
quadratic fashion using the F-statistic
\newline
\newline
\begin{displaymath}
F=\frac{RSS_1-RSS_2}{p_2-p_1} \times \frac{n - p_2}{RSS_2}
\end{displaymath}

where $RSS_i$ is the residual sum of squares for model $i$, $p_i$ is
the number of parameters for model $i$, and $n$ is the number of data
points. To handle the finite size sampling error, the data was binned
using the threshold binning method \cite{adami2002critical}. The
critical value for when aberrant connectivity occurred was found by
determining the local maxima of the F-values \cite{wald43}.

\section{Graph Partitioning Using A Bloom Filter}

We used the Bloom filter data structure containing the k-mers from a
dataset to discover components of the graph, i.e. to partition the
graph.  Here a component is a set of k-mers whose originating reads
overlap transitively by at least $k-1$ base pairs.  Reads belonging
only to small components can be discovered and eliminated in fixed
memory using a simple traversal algorithm that truncates after
discovering more than a given number of novel k-mers.  For discovering
large components we tag the graph at a minimum density by using the
underlying reads as a guide.  We then exhaustively explore the graph
around these tags in order to connect tagged k-mers based on graph
connectivity.  The underlying reads in each component can then be
separated based on their partition.

\section{Assembler software}

We used ABySS v1.3.1 and Velvet v1.1.07 to perform assemblies.  The
ABySS command was: {\tt mpirun -np 8 ABYSS-P -k31 -o contigs.fa
  reads.fa}.  The Velvet commands were: {\tt velveth assem 31 -fasta
  -short reads.fa \&\& velvetg assem}.  We did not use Velvet for the
partitioning analysis because Velvet's error correction algorithm is
stochastic and results in dissimilar assemblies for different read
order.

\section{Software and Software Availability}

We have implemented this compressible graph representation and the
associated partitioning algorithm in a software package named khmer.
It is written in C++ and Python 2.6 and is available under the BSD
open source license at https://github.com/ged-lab/khmer.  The graphviz
software package was used for graph visualizations. The scripts to
generate the figures of this paper are available in the khmer
repository.

\end{materials}

\begin{acknowledgments}

We thank Chris Adami, Qingpeng Zhang, and Tracy Teal for thoughtful
comments, and Jim Cole and Jordan Fish for discussion of future
applications.  In addition, we thank three anonymous reviewers for
their comments, which substantially improved the paper.  This project
was supported by AFRI Competitive Grant no. 2010-65205-20361 from the
USDA NIFA and NSF IOS-0923812, both to CTB.  The MSB2 soil metagenome
was sequenced by the DOE's Joint Genome Institute through the Great
Lakes Bioenergy Research Center (DOE BER DE-FC02-07ER64494).  AH was
supported by NSF Postdoctoral Fellowship Award \#0905961.


\end{acknowledgments}




\end{article}

\begin{figure}
\centering
\includegraphics[width=5in]{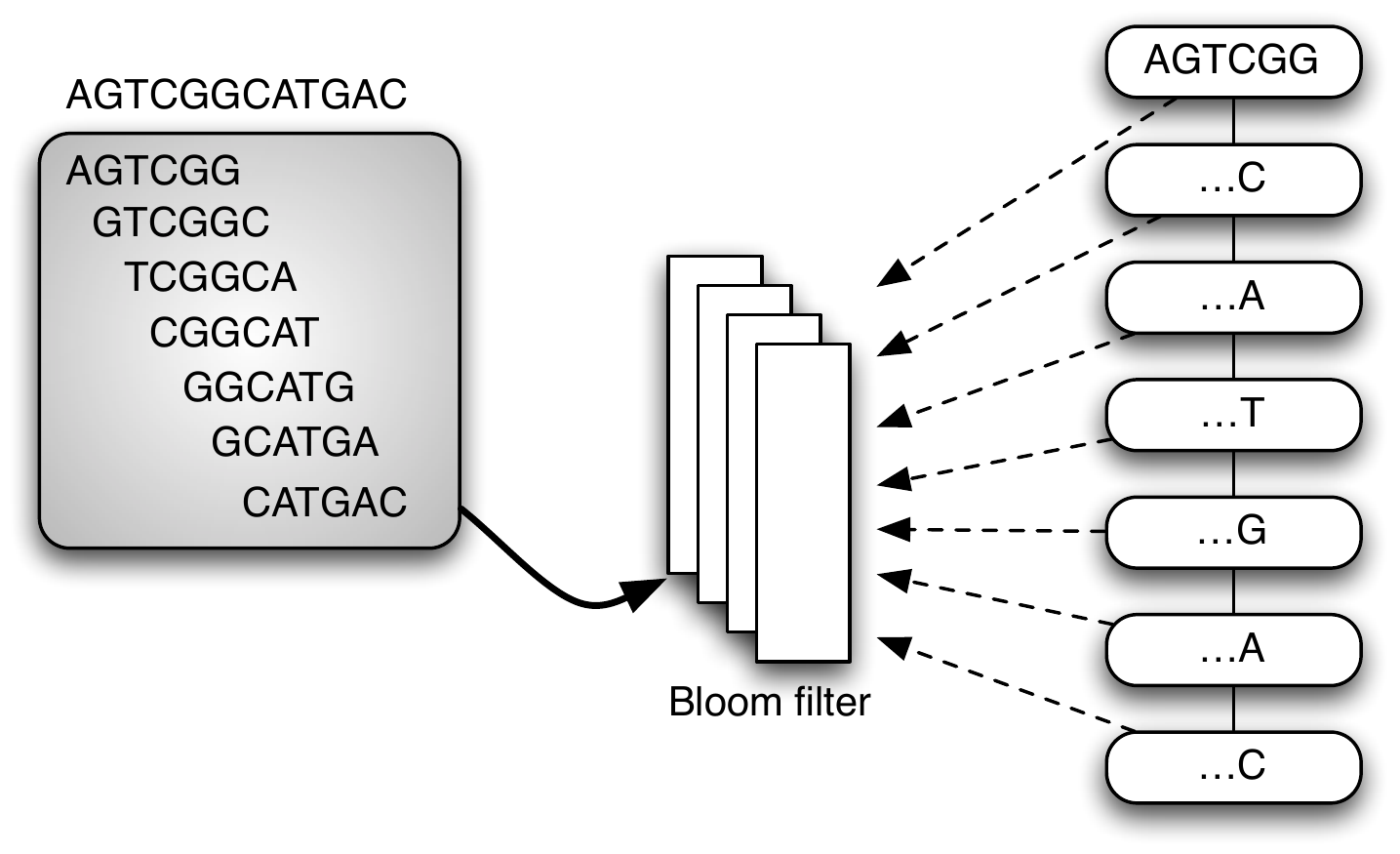}
\caption{Storing de Bruijn graphs in Bloom filters.  Longer sequences (top left) are broken down into k-mers (bottom left; here, k=6) and stored in the Bloom filter (middle).
The graph (right) is traversed by starting at a k-mer and then testing all possible 1-base pre- and post-fixes for presence in the Bloom filter.}

\label{fig:bloomgraph}
\end{figure}

\begin{table*}
\centering
\caption{Bits per k-mer for various false positive rates.}
\begin{tabular*}{\hsize}{@{\extracolsep{\fill}}cccc}
\hline
False positive rate & Bits/k-mer \\ \hline
0.1 \% & 14.35 \cr
1 \% & 9.54 \cr
5 \% & 6.22 \cr
10 \% & 4.78 \cr
15 \% & 3.94 \cr
20 \% & 3.34 \cr
\hline\end{tabular*}
\label{table:bitskmer}
\end{table*}

\begin{table}
\centering

\caption{Effects of loading \emph{E. coli} data at different false positive rates}
\begin{tabular*}{\hsize}{@{\extracolsep{\fill}}ccccccc}
\hline
Graph & Total k-mers & False connected k-mers & \% Real & Deg $> 2$ & Mem (bytes) \\ \hline
\emph{E. coli} at 0\% & 4,530,123 & 0 & 100 & 50,605 & $2.1 \times 10^{9}$ \\
\emph{E. coli} at 1\% & 4,814,050 & 283,927 & 94.1 & 313,844 & $5.4 \times 10^6$ \\
\emph{E. coli} at 5\% & 6,349,301 & 1,819,178 & 71.3 & 1,339,102 & $3.5 \times 10^6$ \\
\emph{E. coli} at 15\% & 31,109,523 & 26,579,400 & 14.6 & 10,522,482 & $2.2 \times 10^6$ \\
Reads at 0\% & 45,566,033 & 41,036,029 & 9.9 & 7,699,309 & $2.1 \times 10^{9}$ \\
Reads at 1\% & 48,182,271 & 43,652,265 & 9.4 & 31,600,208 & $5.4 \times 10^7$ \\
Reads at 5\% & 62,019,545 & 57,489,537 & 7.3 & 42,642,203 & $3.6 \times 10^7$ \\
Reads at 15\% & 231,687,063 & 227,157,037 & 1.9 & 113,489,902 & $2.3 \times 10^7$ \\
\hline
\end{tabular*}
\label{table:ecoli}
\end{table}

\begin{table}
\centering
\caption{Partitioning results on a soil metagenome at k=31.}

\begin{tabular*}{\hsize}{@{\extracolsep{\fill}}cccc}
False positive rate & Total memory use (improvement) & Largest partition size in reads \cr
\hline
1\% & 1.75GB (18.8x) & 344,426 \cr
5\% & 1.20GB (27.5x) & 344,426 \cr
10\% & 0.96GB (34.37x) & 344,426 \cr
15\% & 0.83GB (39.75x) & 344,426 \cr
\hline
\end{tabular*}

\label{table:parts}
\end{table}

\setcounter{figure}{0}
\makeatletter
\renewcommand{\thefigure}{S\@arabic\c@figure}
\begin{figure}
\center{\includegraphics[width=5in]{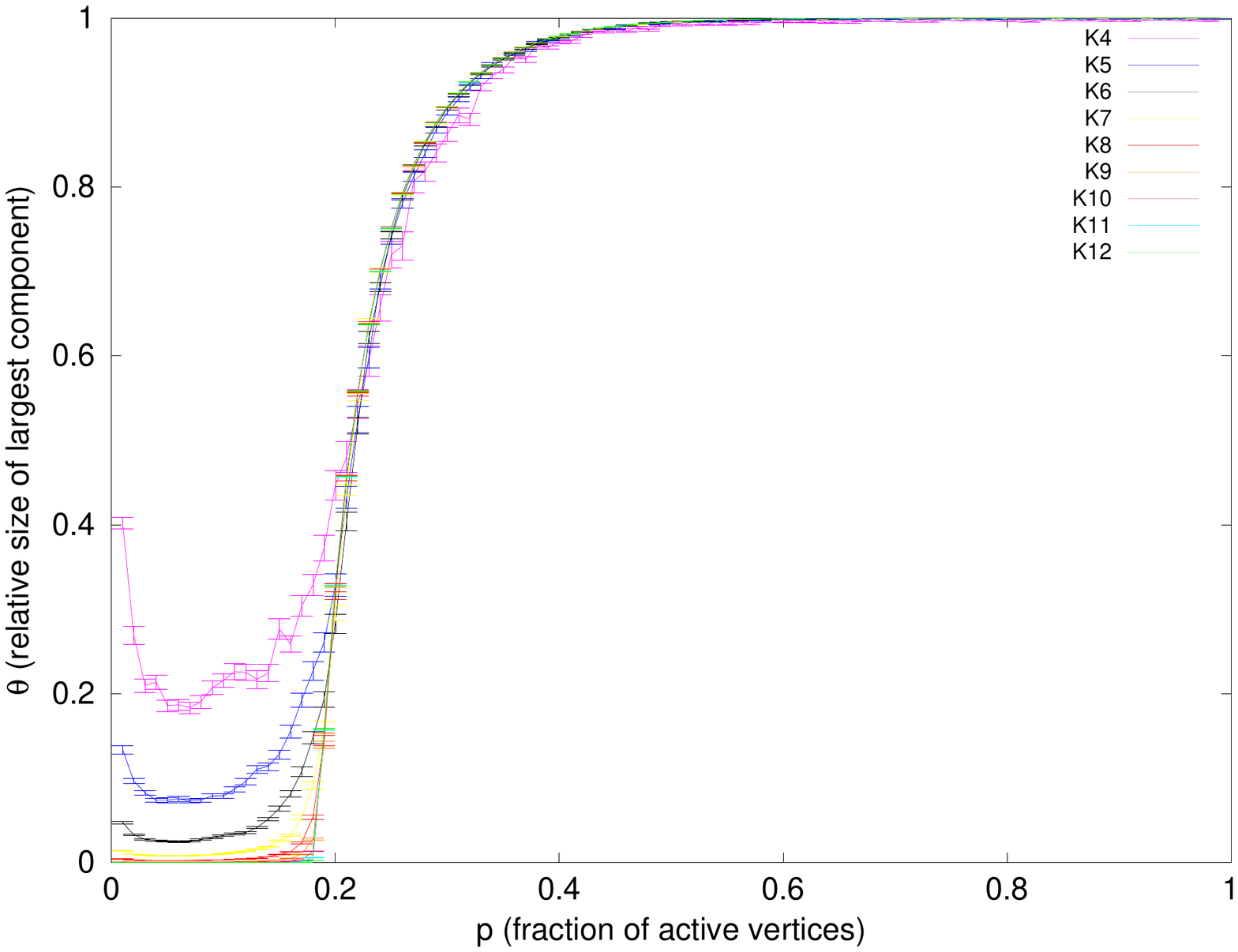}}

\caption{Demonstration of k independence by determining the
  percolation threshold with multiple values of k (5-12).  $p$ on the
  x axis is the fraction of nodes present, and $\theta$ on the y axis
  is the fraction of nodes in the largest component.  Lower values of
  k have greater finite size sampling errors.}

\label{fig:kindependence}
\end{figure}

\begin{figure}
\center{\includegraphics[width=5in]{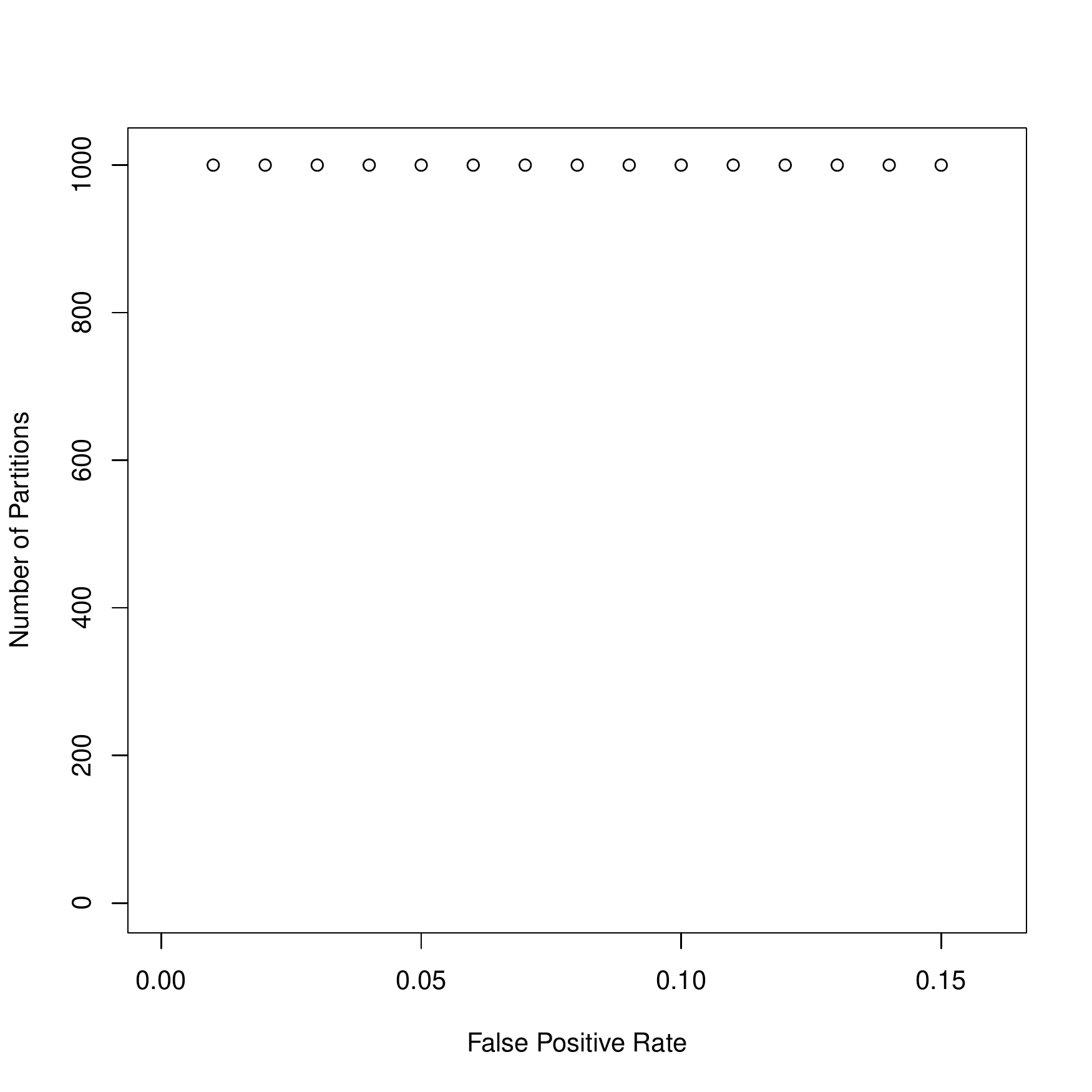}}

\caption{The graph shows the number of partitions for a simulated dataset with
  1,000 contigs of 10,000 bp each (circles). For $n=5$ different
  combinations of hash table sizes, there was no variation in results
  for the simulated dataset.}

\label{fig:partfp}
\end{figure}

\makeatother


\end{document}